\documentclass[twocolumn,aps,floatfix,showpacs,showkeys,prl]{revtex4}
\usepackage{epsfig,color}
\usepackage{graphicx}
\usepackage{amsmath}
\usepackage{amssymb}
\usepackage{amsfonts}
\usepackage{tabularx}
\usepackage{color}
\usepackage{dcolumn}% Align table columns on decimal point
\usepackage{bm}% bold math
\usepackage{floatflt}
\usepackage{setspace}
\usepackage{subfigure}

 \def\cro2{CrO$_2$}
\def\uga3{UGa$_3$}
\def\uge2{UGe$_2$}
 \def\bmao#1{\mbox{\boldmath $#1$}}

\begin{document}
\title{Cancellation of probe effects in measurements of spin polarized momentum density by electron positron
 annihilation.}
\author{M. Biasini$^{1,\#}$, J. Rusz$^{3}$}
\affiliation{1 Department of Physics, University of California at
Riverside, Riverside, California 92521, USA\\
2 Department of Electronic Structures, Charles University, Ke
Karlovu 5, 12116 Prague 2, Czech Republic}

\date{\today}
\begin{abstract}
Measurements of the two dimensional angular correlation of the
electron-positron annihilation radiation have been done in the
past to detect the momentum spin density and the Fermi surface. We
point out that the momentum spin density and the Fermi Surface of
ferromagnetic metals can be revealed within great detail owing to
the large cancellation of the electron-positron matrix elements
which in paramagnetic multiatomic systems plague the
interpretation of the experiments. We prove our conjecture by
calculating the momentum spin density and the Fermi surface of the
half metal \cro2, who has received large attention due to its
possible applications as spintronics material.
 \end{abstract}
 \maketitle

\begin{figure}[btph]

\centerline{\psfig{file=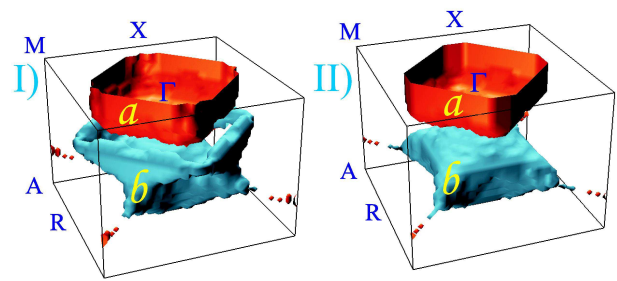,width=8.3cm}}
 \caption{ The two FS sheets (a,b) of FM \cro2, shown in half BZ (Online
 color: a:orange, electronlike; b:blue, holelike) produced with  U$_{xc}$ from  LSDA (I) and GGA (II) \cite{dft,GGA}.
 Here and in the next figures capital letters label the high symmetry points of the BZ.}
 \label{FS_lda_gga}
\end{figure}

To a great extent, the Fermi surface (FS) can regarded as the
defining property of a metal.  It is an ubiquitous concept which
appears in numberless works devoted to the study of the electronic
structure of systems which, under some particular circumstance,
seem to show metallic behaviour.  In several cases, the initial
task of establishing  metallic behavior is not  easy since
different probes can yield contrasting answers.  Clearly,  an
experimental investigation of the FS  implies the observability of
the electrons sitting in the {\em partially filled } energy bands.
which, in turn, requires interaction of the conduction electron
with the experimental probe.   \\
\indent In the case of a measurement of the two dimensional
angular correlation of electron positron annihilation radiation
(2D-ACAR), the FS is revealed through discontinuities (breaks) in
the electron positron momentum density, $ \rho^{ep} (\mathbf{p})$
\cite{berko} at points $\bm{p}_{F_l}= (\bm{k}_{F_l} + \bm{G}) $,
where $ \bm{G} $ is a reciprocal lattice vector and $ \bm{k}_{F_l}
$ are the reduced Fermi wave vectors in the first BZ. While the
locations of the FS breaks are faithfully preserved
\cite{majumdar}, the resulting single particle electron momentum
density, $ \rho^{e} (\mathbf{p})$, is severely modulated by the
non uniform positron (spatial) density and the electron-positron
Coulomb interaction. In this letter we show that in the special
case of a ferromagnetic sample, the probe modulation is strongly
suppressed, allowing a faithful representation of the spin
polarized momentum density and unambiguous interpretation of the
data. The FS breaks are reinforced by the Lock-Crisp-West (LCW)
transformation \cite{lock}, consisting of folding the momentum
distribution $ \rho^{ep} (\mathbf{p}) $ back onto the first BZ by
translation over the appropriate vectors $\mathbf{G}$. The result
of the summation (denoted as LCW density) is \cite{shiotani},
\begin{equation}
\rho_{_{LCW}}^{ep} (\mathbf{k}) = \sum_n \theta (E_F -
\epsilon_{\mathbf{k},n} ) \int | \psi_{\mathbf{k}}^{n}
(\mathbf{r})|^2 \, | \phi \mathbf{(r)}|^2 \, g(\mathbf{r}) d
\mathbf{r}. \label{lcw}
\end{equation}
Here $ \psi_{\mathbf{k}}^n $ and $ \phi$ denote the electron and
positron wave function, respectively, E$_F$ is the Fermi level and
$\epsilon_{\mathbf{k},n}$ is the energy eigenvalue of the electron
from band {\em n} with Bloch wavevector $\mathbf{k}$. The
enhancement factor \cite{boronski} $g(\mathbf{r})$, describes the
enhancement of the electronic density at the positron location due
to the Coulomb force.  If the  $\mathbf{k}, n$ dependence of the
overlap integral in eq.~\ref{lcw} is negligible,
$\rho_{_{LCW}}^{ep} (\mathbf{k}) $ is reduced to the {\em
occupancy}, i.e. the number of occupied bands per $\mathbf{k}$
point. The FS manifolds are the loci of the breaks of the
occupancy.  A significant drawback of 2D-ACAR is that all the
outer electrons overlap (to some extent) with the positron probe.
Therefore, when the orbital character of one full valence band
changes noticeably in the BZ, the related LCW density acquires a
k-dependence which is superimposed to the changes in the occupancy
due to the FS breaks of the conduction bands. For example, our
recent work on \uga3 \cite{uga3} has detected a noticeable change
in the p character of three valence bands located 1-2 eV below
E$_F$ which did obscure greatly the visibility of the FS. Further
notable examples are high-Tc superconductors where strong positron
wave function effects prevented the observation of the critical FS
sheets linked to the Cu-Oxide planes, which are responsible of
superconductivity
\cite{west92,sterne93}.\\
\indent Obviously, the elimination of the contribution of all the
filled bands from the LCW  summation would increase greatly the
visibility of the FS whenever the overlap integral of eq.
\ref{lcw} is \bmao{k}(valence)-dependent. This favorable situation
is indeed realized in the measurement of the spin polarized bands
of ferromagnetic metals. In this case, the employment of 2D-ACAR
experiments
 \cite{berkofm,manuelni,hanssenth} hinges on
 two facts: i) the intrinsic polarization $P_{e^+}$ for positrons produced during $\beta$ decay (for the $^{22}$Na
 source  $P_{e^+}$, averaged over angle and velocity, is about 36\%).
 ii) the annihilation selection rule which requires that the positron may undergo 2$\gamma$ annihilation only if the
spins of the annihilating pair form a singlet state.\\
\indent Therefore, by performing  2D-ACAR measurements with the
magnetic substance polarized (by an external magnetic field) in
directions  respectively parallel and antiparallel to the average
positron polarization (which is unchanged upon reversal of the
magnetic field) will then yield an observable imbalance of
2$\gamma$ annihilations with respect to the majority or minority
spins. The experimental spectra  taken when the positron
polarization is parallel (antiparallel) to the polarizing magnetic
field will be
\begin{eqnarray}
\begin{array}{l}
\rho^{ep}_{par,antipar}(\mathbf{p})=1/2 (1\pm
P_{e^+})[N^M(\mathbf{p})+N^{NM}(\mathbf{p})/2]+   \vspace{0.4cm}\\
\;\;\;\;\;\; + 1/2 (1\mp
P_{e^+})[N^m(\mathbf{p})+N^{NM}(\mathbf{p})/2] ,
\end{array}
\label{magnet0}
\end{eqnarray}
where N$^{NM}$ refers to the contribution from the  spin
degenerate (i.e. non magnetic) bands. Subtraction of
$\rho^{ep}_{par}$ from $\rho^{ep}_{antipar}$ yields the net
momentum spin density $\rho^{ep}_{spin}(\mathbf{p})$.
\begin{equation}
\rho^{ep}_{spin}(\mathbf{p})= P_{e^+} [
\rho^{ep}_{par}(\mathbf{p})- \rho^{ep}_{antipar}(\mathbf{p}) ].
\label{magnet1}
\end{equation}
Equations \ref{magnet0} and \ref{magnet1} are equally applicable
to the electron positron momentum density,
$\rho^{ep}(\mathbf{p})$, or to the LCW density, yielding for the
latter case, the LCW net spin density $
 \rho_{_{LCW} \, spin}^{ep}(\mathbf{k})
 $. \\
\indent  The main point of this work is the disappearance of
N$^{NM}$ and all the positron wave function effects related to the
bands in question.  As obvious as it may appear, this interesting
result has, to our knowledge, never been
pointed out and should prove to be very useful in future studies of ferromagnetic metals. \\
\indent To investigate our conjecture we have chosen to examine
\cro2 which
 is predicted to be a half metal by a variety of ab initio band structure calculations
\cite{sorantin,mazin,kunes,toropova}. Due to this result \cro2 has
attracted much attention in the field of spintronics for its
potential use as injector of a highly (nominally 100\%) spin
polarized current in the future {\em field effect spin
transistor}.  A puzzling result of the calculations is that
different recipes to approximate the exchange correlation
potential, U$_{xc}$ (local spin density approximation (LSDA)
\cite{dft} and Generalized Gradient Approximation (GGA)
\cite{GGA}) lead to unusually noticeable differences in the FS
topology \cite{mazin}.
 A further interesting result is that the standard LSDA and GGA  seem to explain
  ultraviolet photoemission spectroscopy (UPES)
experiments\cite{kurmaev} better than the LDA+$U$ method,
\cite{toropova}, which usually is more suited to treat strong
electron correlations.  \\
\indent In Refs.~\onlinecite{uga3,rusz1} we have presented a
method to calculate directly the LCW density via eq.~\ref{lcw}.
  As a base for our calculation we used the scalar relativistic full-potential
linearized augmented plane waves method (FP-LAPW)
 implemented in the WIEN2k package \cite{wien2k}.  Compared to similar works of other authors
 \cite{pickett,shiotani}, our scheme, including electron-positron correlation effects \cite{boronski}, being {\em full potential} \ and compatible
 with any option of WIEN2k, which include spin-orbit (s.o.) interaction, LDA+$U$ and orbital
 polarization, is more suited  to the study of narrow bands and electron
 correlations.\\
 \indent  The extension of the scheme to spin polarized
calculations is
 relatively straightforward.  The same steps used in the
 paramagnetic calculations are done for spin up and spin down
 electrons separately.  In the absence of s.o. interaction the result consists of {\em
 pure} up spin and down spin bands.
  After the inclusion of s.o.\ each Bloch state becomes a combination of
spin up and down components. For each  band crossing E$_F$ there
is now an up and down spin submanifold, both having the same
occupancy (and therefore the same Fermi surface). The amplitudes
of these submanifolds are, however, different. Owing to parity and
angular momentum conservation, the positron essentially {\em
projects} the electronic states over the spin up or down
manifolds. \\
\indent The crystal structure of \cro2 is simple tetragonal
(Rutile type, space group P4$_2/mnm$, \cite{sorantin}) with two
formula units per unit cell.  Details of the calculations,
performed
 adopting U$_{xc}$ according to GGA and LDA
 in the paramagnetic  and
ferromagnetic phases, are reported elsewhere \cite{cro2after}. The
calculated majority spin bands  are in excellent agreement with
those reported by Mazin et al \cite{mazin}. The main finding of
the calculation is the half metallic behaviour, with a large
energy gap ($\simeq$2eV) in the minority spin bands. The
conduction bands have a very strong Cr $d$ character, which is
ascribed to states of {\em t$_{2g}$} symmetry. It is worth noting
that the Cr-d orbital character of all the conduction bands is
rather constant along each of the high symmetry directions.
\\
\indent As mentioned above, the topology of the holelike FS sheet,
shown in Fig. \ref{FS_lda_gga}(I,II)b depends noticeably on the
kind of U$_{xc}$ (LSDA,GGA).
% \vspace{-.5cm}
  Whereas in the GGA case
this FS has a simple, pillow-like shape
(Fig.~\ref{FS_lda_gga}/II-b), the corresponding FS manifold
resulting from LSDA (Fig.~\ref{FS_lda_gga}/I-b) is a pseudosphere
connected to a toroidal structure with rhombic shape.  On the
other hand, the electronlike, $\Gamma$ centred structure,
 yielded by the two
calculations is  very similar.
 Since \cro2 is a compensated metal (a necessary condition of any half
 metal),  the electronlike and
holelike sheets have equal volume.  The Fermi volumes resulting
from LSDA and GGA differ slightly, corresponding to 12.8\% of the
BZ and 10.7\% of the BZ for LSDA and GGA, respectively.  \\
\begin{figure}[btph]
%\vspace{-0.5cm}
\centerline{\psfig{file=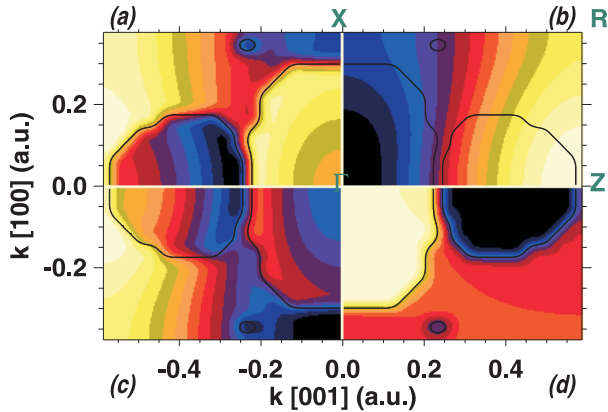,width=8.cm}} \caption{(Online
color.) FLAPW calculations for \cro2  in the (010) plane of the
tetragonal BZ: (a) up spin LCW density; (b) down spin LCW density;
(c) (a)+(b); (d) (a)-(b). The black contour denote the
intersections of the two main FS sheets (see fig \ref{FS_lda_gga})
with the (010) plane. Note that the orizontal direction is along
[001]} \label{c100}
\end{figure}
 The predicted response of a 2D-ACAR experiment experiment can be
elucidated by a slice of the calculated LCW density (adopting
LSDA) in the (010) plane (GGA yields little difference in this
plane; recall Fig \ref{FS_lda_gga}). In Fig.~\ref{c100} panels (a)
and (b) refer to the majority and minority spins LCW densities,
respectively, whereas panels (c) and (d) denote sum and difference
of panels (b) and (b). These panels would be obtained by an
experiment with 100\% spin polarized positrons impinging a sample
where the polarizing magnetic field is parallel or antiparallel to
the positron polarization. In panel (a) the breaks pertaining to
the electronlike and holelike FS sheets appear rather clearly
(compare the changes of gray scale (or color) with the contour
line marking the FS breaks), in spite of a strong modulation
caused by a non uniform positron density. This modulation is the
only feature appearing in panel (b), consistent with a negligible
spin up-down mixing due to s.o. (WIEN2k yields 99.97\% electron
polarization at E$_F$). In panel (c) we have summed panels (a) and
(b) to notice that positron wavefunction effects are reinforced at
the expense of the FS signatures.

\begin{figure}[btph]
% \vspace{-.3cm}
\centerline{\psfig{file=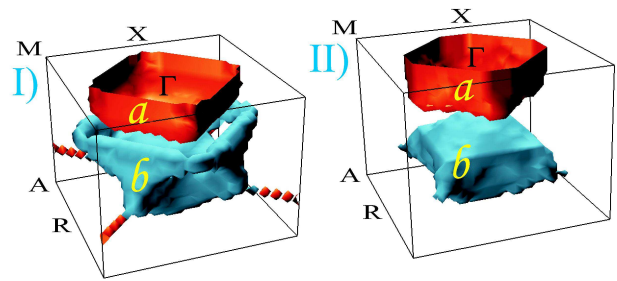,width=8.3cm}}
 \caption{ Spin polarized FSs seen by 2D-ACAR experiments in \cro2,  shown in half BZ (theoretical prediction).
  The FSs  are identified by the two isodensity surfaces (a,b)
 extracted from $\rho^{ep}_{{_{LCW}} \ spin}(\mathbf{k})$
 as described in the text.  (Online
 color: a:orange, electronlike; b:blue, holelike). The pertaining  LCW densities were
   produced with U$_{xc}$ from  LSDA (left, I) and GGA (right, II) \cite{dft,GGA}.
}
 \label{FS_diff}
\end{figure}
 %\vspace{-.5cm}
The very appealing message, and the main point of this paper, is
however provided by panel (d), where most of the positron wave
function
 effects cancel out and the almost intact topology of the FS is restored.

As anticipated,  the difference eliminates the contribution from
the non spin polarized bands and any  positron wave function
effect to be ascribed to them. The cancellation of positron
effects caused by all the non magnetic bands, which is obviously
effective also for ferromagnetic insulators \cite{chiba}, is
clearly of great importance in systems with several atoms in the
unit cell and, consequently, several bands near to $E_F$. A
similar effect can be inferred from Fig. 9 of
Ref.\cite{hanssenth}a), referring to the half metal NiMnSb. In
\cro2 the visibility of the FS is further enhanced by the weakness
of the positron modulation of the conduction bands. This result is
consistent with the constancy of the d orbital character in the
BZ for all the conduction bands, noted above. \\
\indent The similarity of panel \ref{c100}(d) with the theoretical
occupancy (whose breaks denote the FS) makes feasible algorithms
aimed at extracting the FS from isodensity surfaces of the LCW
density. This task would be impossible for the  data set shown in
panel \ref{c100}(a), where, owing to positron modulation, the
amplitude of the LCW density nearby the Fermi break of a single
sheet changes noticeably in the BZ. Conversely,
$\rho^{ep}_{{_{LCW}} \ spin}(\mathbf{k})$ presented in panel
\ref{c100}(d) allows to separate very clearly the two conduction
bands and apply the method presented  in Ref. \cite{biaerga} which
identifies a multi-sheet FS in terms of isodensity surfaces
selected at the loci of maximal   amplitude variation of the LCW
density. It is worth noticing that in the half metals  the
analysis of $\rho^{ep}_{{_{LCW}} \ spin}(\mathbf{k})$ for the
search of the FS manifolds is particularly suited.  In fact, since
the subtraction of the {\em insulating} LCW (down) spin density
from the {\em conducting}  LCW (up) spin density preserves the
proper sign in the jumps of the occupancy due to the FS breaks,
maxima of  $\rho^{ep}_{{_{LCW}} \ spin}(\mathbf{k})$ will
correspond to maxima of the occupancy and viceversa. Fig
\ref{FS_diff} shows the resulting isodensity surfaces of
$\rho^{ep}_{{_{LCW}} \ spin}(\mathbf{k})$. The similarity with the
true FS shown in Fig. \ref{FS_lda_gga} is striking.
\begin{figure}[btph]
% \vspace{-0.3cm}
\centerline{\psfig{file=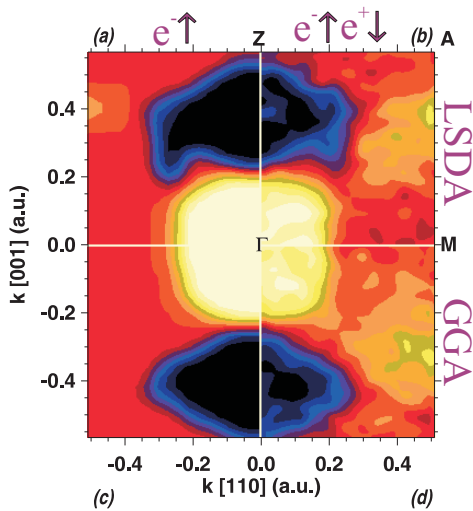,width=6.2cm}} \caption{(Online
color.) FLAPW calculations for \cro2 in the (110) plane of the
tetragonal BZ. (a) LSDA up spin occupancy convoluted with the
experimental resolution; (b) LSDA simulated experimental
difference spectra (inclusive of statistical noise; see text),
$\rho_{_{LCW} \, spin}^{ep}(\mathbf{k})$; (c) same as (a) for GGA;
(d) same as (b) for GGA.} \label{c110}
\end{figure}
The cancellation of positron wave function effects shows that
2D-ACAR has the power to investigate the FS predictions of LSDA
and GGA, shown in Fig.~\ref{FS_lda_gga}. The difference in the FS
is well revealed in the (110) plane. To establish to what extent
feasible experiments can accomplish this task, in Fig.~\ref{c110}
we have simulated in full the output of the experiment, according
to eq.~\ref{magnet0}. The steps adopted to produce  Fig.
\ref{c110} are, in succession: i) Calculation (adopting U$_{xc}$
from GGA and LSDA) of $\rho_{_{LCW} \, par}^{ep}(\mathbf{k})$ and
$\rho_{_{LCW} \, antipar}^{ep}(\mathbf{k})$ according to the
mixing shown in eq.~\ref{magnet0}, employing the realistic
positron polarization $P_{e^+}=0.35$; ii) Convolution of
$\rho_{_{LCW} \, par}^{ep}(\mathbf{k})$ and $\rho_{_{LCW} \,
antipar}^{ep}(\mathbf{k})$ with the experimental resolution $R$
\cite{resol}; iii) Perturbation of the two simulated densities
with statistical noise. It is assumed that the LCW density was
reconstructed from 5 projections, each collecting 5$\times 10^8$
coincidence counts, and that the spectra were symmetrized in the
3D reconstruction procedure \cite{biaerga}; iv) Construction of
$\rho_{_{LCW} \,
spin}^{ep}(\mathbf{k})$. \\
\indent  Figure~\ref{c110} shows the electronic occupancy
(reflecting the FS topology) convoluted with the experimental
resolution [panels (a),(c)] compared to the difference spectra
resulting from steps i)-iv) [panels (b),(d)] for  LSDA and GGA,
respectively. Interestingly, the bulging of the LSDA holelike FS
at $\simeq$ (0.27,0.20) a.u.[panel (a)], absent in the GGA
holelike FS [panel (c)] is well reproduced by the difference
$\rho_{_{LCW} \, spin}^{ep}(\mathbf{k}) $ [panel (b)] and not
present in panel (d).  Obviously, to reveal the FS subtleties at
stake, arrays with large number of counts are required,
particularly when differences between experimental spectra are
analyzed. With this caveat, we predict that a 2D-ACAR experiment
should decide over this issue.  \\
\indent In conclusion,  we have pointed out the power of the
magnetic LCW procedure to eliminate large part of the modulation
of the LCW density due to the non uniform positron density and
reveal the FS in great detail. We have proved that this indeed
happens by applying our scheme to calculate the LCW density of the
half metal \cro2. We have simulated in full the output of an
experiment performed with partial polarization of the positron,
dictating the amount of data collection required to have
appropriate signal to noise ratios.  Note that the ability to
include s.o. effect can be employed to corroborate 2D-ACAR
measurements aimed at determining the polarization of the
conduction electrons at the Fermi surface. This type of
information could be of critical importance for the design of
novel materials to be employed  in the implementation of the
future
spin-based FET. \\
 \indent We thank Allen Mills for stimulating discussions. Work supported by CNID and
project MSM 0021620834 financed by Ministry of Education of Czech Republic.


\begin{thebibliography}{99}
\bibitem{resol}
The adopted value $R=0.0685$~a.u., corresponding to 9\%
 of the BZ along the [100] direction, is typical of existing 2D-ACAR setups.
\end{thebibliography}
\end{document}